\documentclass[prl,aps]{revtex4} 

\usepackage{graphicx}
\usepackage{dcolumn}
\usepackage{bm}

\begin{document}

\title{Nonextensive statistical mechanics and complex scale-free networks}
\author{Stefan Thurner}
\affiliation{Complex Systems Research Group  HNO, Medical University of Vienna, W\"ahringer G\"urtel 18-20, A-1090 Vienna, Austria }

\maketitle

One explanation for the impressive recent  boom in network theory 
might be that it provides a promising tool for 
an understanding of complex systems. Network theory 
is mainly focusing on discrete large-scale topological structures  
rather than on microscopic details of interactions of its elements. 
This viewpoint allows to naturally treat collective phenomena 
which are often an integral part of complex systems, such as 
biological or socio-economical phenomena.
Much of the attraction of network theory arises  from 
the discovery that many networks, natural or man-made, seem to exhibit some 
sort of  universality, meaning that most of them belong to  
one of three classes: {\it random}, {\it scale-free} and {\it small-world} 
networks. 
Maybe most important however for the physics community is, that due to its 
conceptually intuitive nature, network theory seems to be within reach 
of a full and coherent understanding from first principles.

Networks are discrete objects made up of a set of nodes which are 
joint by a set of links. If a given set of $N$ nodes is linked by a fixed 
number of links in a completely random manner, the result
is a so-called {\it random network}, whose characteristics can be rather 
easily understood.
One of the simplest measures describing a network in statistical 
terms is its degree distribution, $p(k)$, (see box 1). In the case of 
random networks the degree distribution is a Poissonian, i.e., the
probability (density) that a randomly chosen node has degree $k$ is given by
$p(k)=\frac{\lambda^k e^{-\lambda}}{k!}$, where $\lambda=\bar k$ is the average 
degree of all nodes in the network.
However, as soon as more 
complicated rules for wiring or growing of a network are considered, 
the seemingly simple concept of a network can become rather involved.
In particular, in many cases the degree distribution 
becomes a power-law, bare of any characteristic scale, which 
raises associations to critical phenomena and scaling phenomena 
in complex systems. This is the reason why these types of networks 
are often called {\it complex networks}. 
A further intriguing aspect of dynamical complex networks is that they 
can potentially provide some sort of toy-model for 'nonergodic' systems, in 
the sense that not all possible states (configurations) are equally 
probable or homogeneously populated, and thus can violate a key assumption 
for systems described by classical statistical mechanics. 

Over the past two decades the concept of {\it nonextensive statistical 
mechanics} has been extremely successful in addressing critical phenomena, 
complex systems and nonergodic systems \cite{tsallis88,gellmann}. 
Nonextensive statistical mechanics  
is a generalization of Boltzmann-Gibbs statistical mechanics, 
where entropy is defined as
\begin{equation}
S_q \equiv \frac{1-\int_0^\infty dk \, [p(k)]^q}{q-1} 
\quad {\rm with \,\, the}\,\, q=1  \,\,{\rm limit} \quad 
S_1=S_{BG} \equiv -\int_0^\infty dk \, p(k) \ln p(k) 
\end{equation}
where $BG$ stands for {\it Boltzmann-Gibbs}. 
If -- in the philosophy of the maximum entropy principle -- 
one extremizes $S_q$ under certain constraints,
the corresponding distribution is the $q$-exponential 
(see box in editorial paper by C. Tsallis and J.P. Boon).
Another sign for the importance and ubiquity of $q$-exponentials in nature 
might be due to the fact that the most general Boltzmanfactor for canonical 
ensembles ({\it extensive}) is the $q$-exponential, as was recently proved in 
\cite{hanel}.
Given the above characteristics of networks and the fact that 
a vast number of real-world and model networks show asymptotic power-law 
degree distributions, it seems almost obvious to look for a 
connection between networks and nonextensive statistical physics. 
                                                                                 
Since the very beginning of the recent modeling efforts of complex networks  
it has been noticed that degree distributions asymptotically 
follow power-laws \cite{albert99}, or even exactly  $q$-exponentials  \cite{albert00}.
The model in \cite{albert99} describes growing networks with a so-called 
preferential attachment rule, meaning that any new node being added to the system 
links itself to an already existing node $i$ in the network with a probability 
that is proportional to the degree of node $i$, i.e. 
$p_{\rm link}\propto k_i$. In  \cite{albert00} this model 
was extended to also allow for preferential rewiring. The analytical solution to the 
model has a $q$-exponential as a result, with the nonextensivity parameter $q$ 
being fixed uniquely by the model parameters.   

However, it has been found that networks exhibiting degree distributions 
compatible with $q$-exponentials are not at all limited to growing and 
preferentially organizing networks.
Degree distributions of real-world networks as well as of 
models of various kinds seem to exhibit a universality in this respect. 
In the remainder we will review a small portion of the variety of networks which potentially
have a natural link to non-extensive statistics. Even though 
there exists no complete theory yet, there is substantial evidence for 
a deep connection of complex networks with the $q \neq 1$ 
instance in nonextensive statistical mechanics.

Recently in \cite{soares} preferential attachment networks have been embedded 
in Euclidean space, where the attachment probability for a newly added 
node is not only proportional to the degrees of existing nodes, but also 
depends on the Euclidean distance between nodes. The model is realized 
by setting the linking probability of a new node to an existing  node 
$i$ to be $p_{\rm link}\propto k_i/r_i^{\alpha}$ $(\alpha\geq0)$, with $r_i$ being the distance 
between the new node and node $i$; 
$\alpha=0$ corresponds to the 
model in \cite{albert99} which has no metrics. In a careful analysis 
the degree distributions of the resulting networks have been shown to 
be $q$-exponentials with a clear $\alpha$-dependence of the nonextensivity 
parameter $q$. In the large $\alpha$ limit, $q$ approaches one, i.e., random networks 
are recovered in the Boltzmann-Gibbs limit. 

In an effort to understand the evolution of socio-economic 
networks, in \cite{white} a model was proposed that builds upon  
\cite{albert00} but introduces  a rewiring scheme which 
depends on the {\it internal} network distance between two nodes, 
i.e., the number of steps needed to connect the two nodes.
The emerging degree distributions have been subjected 
to a statistical analysis where the hypothesis of $q$-exponentials 
could not be rejected. 

A model for nongrowing networks which was recently put forward in 
\cite{thutsall} also unambiguously exhibit $q$-exponential degree distributions. 
This model was motivated by interpreting networks as a certain type 
of 'gas' where upon an (inelastic) collision of two nodes,  links get 
transfered in analogy to the energy-momentum transfer in real gases. 
In this model a fixed number of nodes 
in an (undirected) network can 'merge', i.e., two nodes fuse into one single 
node, which keeps the union of links of the two original nodes; 
the link connecting the two nodes before the merger is removed.
At the same time a new node is introduced to the system and is linked 
randomly to any of the existing nodes in the network \cite{sneppen}.  
Due to the nature of this model the number of links is not strictly conserved -- 
which can be thought of as jumps between discrete states in some 'phase space'.
The model has been further generalized to exhibit a distance dependence as in 
\cite{soares}, however $r_i$ not being Euclidean but internal distance. 
Again, the resulting degree distributions have $q$-exponential form. 
In Fig. 1 we show a snapshot of this type of network
{\it pars pro toto} for the many models exhibiting $q$-exponential degree distributions.
The corresponding (cumulative) degree distribution is shown in Fig. 2 in log-log scale,
clearly exhibiting a power-law. 
Figure 3 shows $q$-logarithms of the degree distribution for several values 
of $q$. It is clear from the correlation coefficient of the $q$-logarithm 
with straight lines (inset) 
that there exists an optimal value of $q$, which makes the $q$-logarithm a 
linear function in $k$, showing that the degree distribution is a $q$-exponential.

A quite different approach was taken in \cite{abe05} where 
an ensemble interpretation of random networks has been 
adopted, motivated by superstatistics \cite{beck}.  
Here it was assumed that the average 
connectivity $\lambda$ in random networks is fluctuating according to 
a distribution $\Pi(\lambda)$, which is sometimes associated with a 
'hidden-variable' distribution. In this sense a network with any degree distribution 
can be seen as a 'superposition' of random networks with the 
degree distribution given by 
$p(k)= \int_0^{\infty} d\lambda \, \Pi(\lambda) \frac{\lambda^k e^{-\lambda}}{k!}$. 
In \cite{abe05} it was shown as an exact example, that a power-law functional form of $\Pi(\lambda)$
leads to degree distributions of Zipf-Mandelbrot form, $p(k) \propto \frac{1}{(k_0+k)^{\gamma}}$, which is 
equivalent to a $q$-exponential with an argument of $k/\kappa$ and given   
the substitutions, $\kappa= (1-q)k_0$ and $q= 1+1/\gamma$. 
Very recently  a possible connection between {\it small-world}
networks and the maximum Tsallis-entropy principle, as well as to the hidden variable method \cite{abe05}, 
has been noticed in \cite{hasegawa}. 

In yet another view of networks from a physicist's perspective, 
networks have recently been treated as statistical systems 
on a Hamiltonian basis.
It has been shown that these systems show a phase transition like behavior
\cite{vicsek}, along which networks structure changes. In the 
low temperature phase one finds networks of 'star' type, meaning that a few nodes are 
extremely well connected resulting even in a discontinuous $p(k)$; in the 
high temperature phase one finds random networks. Surprisingly, for a special  
type of Hamiltonians networks with $q$-exponential degree distributions 
emerge right in the vicinity of  the transition point \cite{biely}.

While a full theory of how complex networks are connected to $q \neq 1$ statistical 
mechanics is still missing, it is almost clear that such a relation should exist. 
It would not be surprising if an understanding of this relation would arise from   
the very nature of networks, being {\it discrete} objects. 
More specifically, it has been  conjectured for nonextensive systems  that the 
microscopic dynamics does not fill or cover the space of states 
(e.g. $\Gamma$-space ($6N$ dimensional phase space) for Hamiltonian systems) 
in a homogeneous and equi-probable manner \cite{gellmann}. 
This  possibly makes phase space for nonextensive systems look like 
a network itself, in the sense that in a network not all possible positions in 
space can be taken, but that microscopic dynamics is restricted onto nodes and 
links. In this view the basis of nonextensive systems could be connected to a 
network like structure of their 'phase space'. 
It would be fantastic if further understanding of network theory could 
propel a deep understanding of nonextensive statistical physics, 
and vice versa, making them co-evolving theories.\\

BOX 1: Some network measures\\

The degree $k_i$ of a particular node $i$ of the network  
is the number of links associated with it. If links are 
directed they either emerge or end at a node, yielding the diction of  
out- or in-degree, respectively. 
The degree distribution $p(k)$ is the probability for finding a node 
with degree $k$ in the network. 
In (unweighted) networks the degree distribution is 
discrete and often reads, $p(k)= p(1) e_{q}^{k/ \kappa}$
with $\kappa>0$ being  some characteristic number of links.
Apart from the degree distribution, important measures to characterize 
network topology are the clustering coefficient $c_i$, and the neighbor 
connectivity $k_{nn}$. The clustering coefficient 
measures the probability that two neighboring nodes of a node $i$  
are also neighbors of each other, and is thus a measure of 
cliquishness within networks. The neighbor connectivity 
is the average degree of all the nearest neighbors of node $i$. 
When plotted as a function of $k$, a nontrivial distribution of 
the average of $c$ 
allows statements about hierarchic structures within the network, 
while $k_{nn}$ serves as a measure of assortativity.  
\\



\newpage

\begin{figure}[htb]
\begin{center}
\begin{tabular}{c} 
\includegraphics[height=230mm]{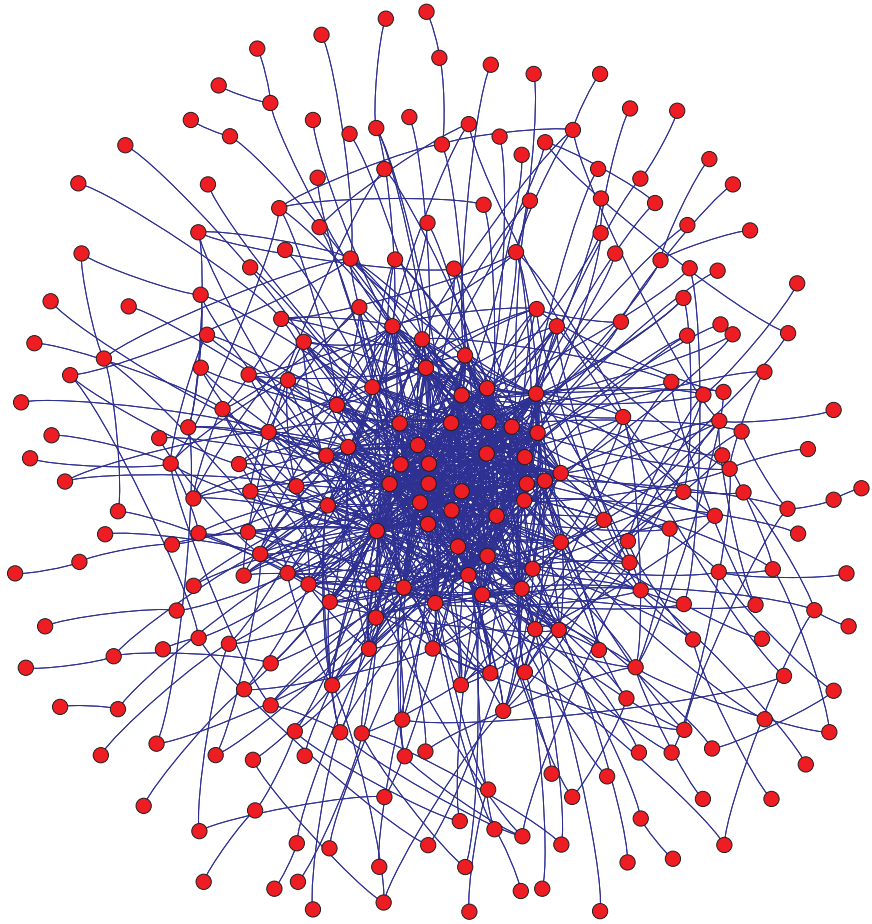} \\
\end{tabular}
\end{center}
\vspace{-3cm}
\caption{Snapshot of a non-growing dynamic network  with 
$q$-exponential degree distribution for $N=256$ nodes and a linking rate of $\bar r=1$, 
for details see \cite{sneppen,thutsall}.
The shown network is small to make connection patterns visible.}
\end{figure}

\begin{figure}[htb]
\begin{center}
\begin{tabular}{c} 
\includegraphics[height=130mm]{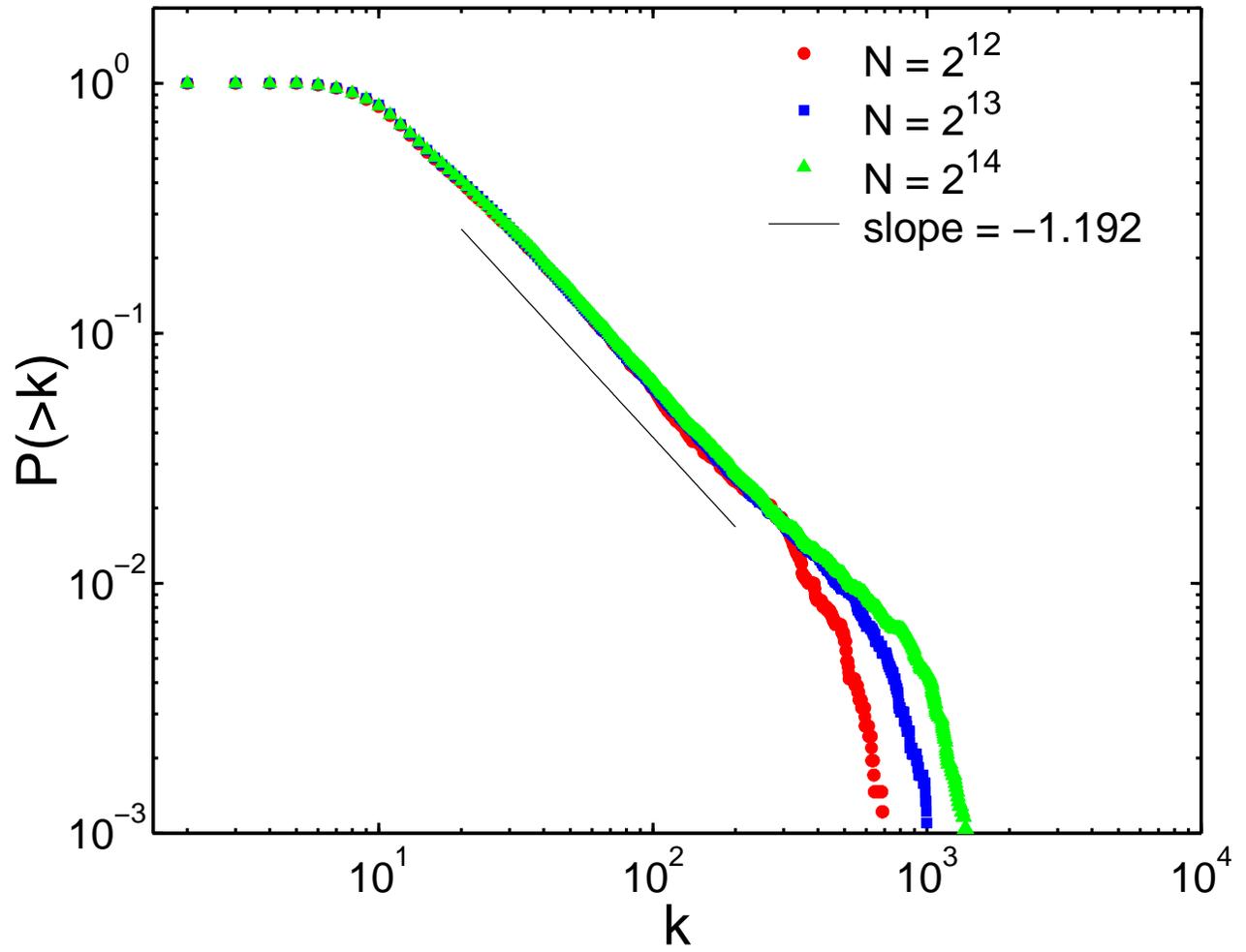} \\
\end{tabular}
\end{center}
\caption{Log-log representation of (cumulative) degree distributions  
for the same type of network with various system sizes $N$ and  
$\bar r =8$.
The distribution functions  are from individual network configurations without averaging over 
identical realizations. On the right side the typical exponential finite size effect  
is seen.}
\end{figure}

\begin{figure}[htb]
\begin{center}
\begin{tabular}{c} 
\includegraphics[height=130mm]{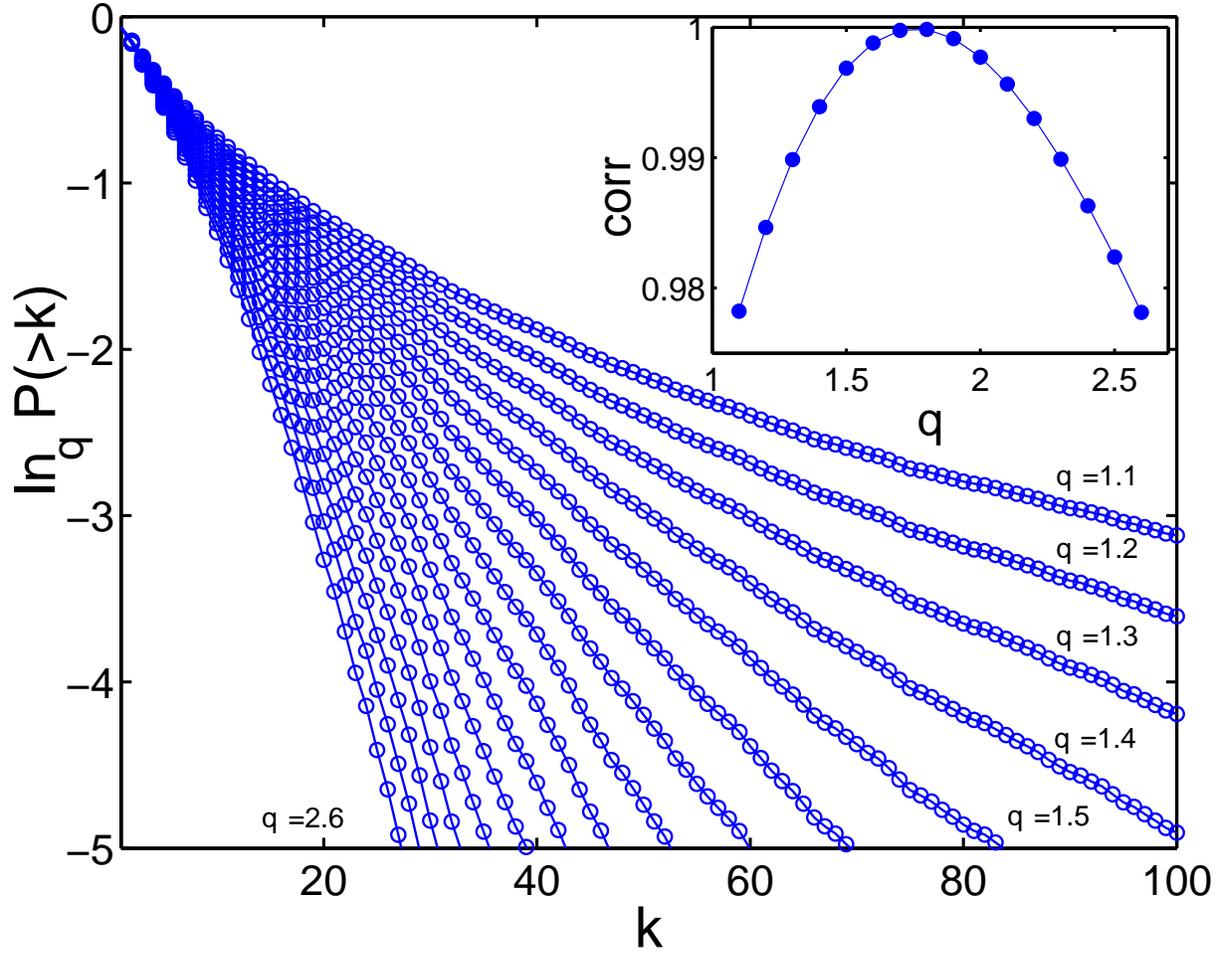}
\end{tabular}
\end{center}
\caption{$q$-logarithm of the (cumulative) distribution function 
from the previous figure as a function of degree $k$. Clearly, 
there exists an optimum $q$ which allows for an optimal linear fit.  
{\it Inset:} Linear correlation coefficient  of 
$\ln_q\, P(\geq k)$ and straight lines for various values of $q$. The optimum
value of $q$ is obtained when $\ln_q\, P(\geq k)$ is optimally linear, i.e., where the 
correlation coefficient has a maximum. A linear  $\ln_q$ means that the 
distribution function is a $q$-exponential; the slope of the linear 
function determines $\kappa$. In this example we get for the optimum $q=1.84$, 
which corresponds to the slope $\gamma=1.19$ in the previous plot.}
\end{figure}

\end{document}